%% file: main.tex
\documentclass[sigconf]{acmart}
\AtBeginDocument{%
  \providecommand\BibTeX{{%
    \normalfont B\kern-0.5em{\scshape i\kern-0.25em b}\kern-0.8em\TeX}}}

\usepackage{booktabs} 

\usepackage[english]{babel}
\usepackage{moresize}
\usepackage{amsmath}
\usepackage{algorithmic}
\usepackage{balance}
\usepackage{comment}
\usepackage{paralist}
\usepackage{bm}
\usepackage{pgfplots}
\usetikzlibrary{pgfplots.dateplot}

\usepackage{flushend}
\usepackage[english]{babel}
\usepackage[latin1]{inputenc}
\usepackage{mathrsfs}
\usepackage{graphicx}

\usepackage{amssymb}
\usepackage{amsfonts}
\usepackage{url}
\usepackage{longtable}
\usepackage{rotating}
\usepackage{multirow}
\usepackage{mathrsfs}
\usepackage{subfigure}
\usepackage{enumitem}
\usepackage[linesnumbered,algoruled,boxed,lined]{algorithm2e}
\usepackage{adjustbox}
\usepackage{hyperref}
\usepackage{pgfplots}
\usetikzlibrary{pgfplots.dateplot}
\usepackage{filecontents}
\definecolor{tblue}{RGB}{31,119,180}
\definecolor{torange}{RGB}{255,127,14}
\definecolor{tgreen}{RGB}{44,160,44}
\definecolor{tred}{RGB}{214,39,40}
\definecolor{tpurple}{RGB}{148,103,189}

\newcommand{\hide}[1]{} 

\newcommand{\ie}{\textit{i}.\textit{e}.}
\newcommand{\eg}{\textit{e}.\textit{g}.} 
\newcommand{\wrt}{\textit{w}.\textit{r}.\textit{t}}

\def\model{GFormer}

\copyrightyear{2023}
\acmYear{2023}
\setcopyright{acmlicensed}\acmConference[SIGIR '23]{Proceedings of the 46th International ACM SIGIR Conference on Research and Development in Information Retrieval}{July 23--27, 2023}{Taipei, Taiwan}
\acmBooktitle{Proceedings of the 46th International ACM SIGIR Conference on Research and Development in Information Retrieval (SIGIR '23), July 23--27, 2023, Taipei, Taiwan}
\acmPrice{15.00}
\acmDOI{10.1145/3539618.3591723}
\acmISBN{978-1-4503-9408-6/23/07}

\begin{document}



\title{Graph Transformer for Recommendation}

\author{Chaoliu Li}
\email{chaoliuli66@gmail.com}
\affiliation{%
  \institution{South China University of Technology}
  \city{Guangzhou}
  \country{China}
}

\author{Lianghao Xia}
\email{aka_xia@foxmail.com}
\affiliation{%
  \institution{University of Hong Kong}
  \city{Hong Kong SAR}
  \country{China}
}

\author{Xubin Ren}
\email{xubinrencs@gmail.com}
\affiliation{%
  \institution{University of Hong Kong}
  \city{Hong Kong SAR}
  \country{China}
  }

\author{Yaowen Ye}
\email{elwin@connect.hku.hk}
\orcid{0009-0006-7227-2306}
\affiliation{%
  \institution{University of Hong Kong}
  \city{Hong Kong SAR}
  \country{China}
}

\author{Yong Xu}
\email{yxu@scut.edu.cn}
\orcid{0000-0001-7183-3155}
\affiliation{%
  \institution{South China University of Technology}
  \city{Guangzhou}
  \country{China}
}

\author{Chao Huang}
\authornote{Chao Huang is the corresponding author. This work was completed when Chaoliu Li was a research intern under the supervision of Chao Huang.}
\email{chaohuang75@gmail.com}
\affiliation{%
  \institution{University of Hong Kong}
  \city{Hong Kong SAR}
  \country{China}
}

\renewcommand{\shortauthors}{Chaoliu Li, Lianghao Xia, Xubin Ren, Yaowen Ye, Yong Xu, \& Chao Huang}


\begin{abstract}
This paper presents a novel approach to representation learning in recommender systems by integrating generative self-supervised learning with graph transformer architecture. We highlight the importance of high-quality data augmentation with relevant self-supervised pretext tasks for improving performance. Towards this end, we propose a new approach that automates the self-supervision augmentation process through a rationale-aware generative SSL that distills informative user-item interaction patterns. The proposed recommender with \underline{G}raph Trans\underline{Former} (\model) that offers parameterized collaborative rationale discovery for selective augmentation while preserving global-aware user-item relationships. In \model, we allow the rationale-aware SSL to inspire graph collaborative filtering with task-adaptive invariant rationalization in graph transformer. The experimental results reveal that our \model\ has the capability to consistently improve the performance over baselines on different datasets. Several in-depth experiments further investigate the invariant rationale-aware augmentation from various aspects. The source code for this work is publicly available at: \url{https://github.com/HKUDS/GFormer}.

\end{abstract}



\maketitle
\input{intro}
\input{relate}
\input{solution}

\input{eval}
\input{conclusion}

\clearpage
\balance
\bibliographystyle{ACM-Reference-Format}
\bibliography{sample-base}


\end{document}

%% file: intro.tex
\section{Introduction}
\label{sec:intro}


Self-supervised learning (SSL) has become a popular solution for addressing the label scarcity issue in recommender systems by generating auxiliary supervision signals from unlabeled data~\cite{zhou2020s3,wei2023multi}. By integrating with graph neural network (GNN) architecture for collaborative filtering, SSL-enhanced graph augmentation has proven effective in modeling user-item interactions with limited training labels. Among contemporary methods, graph contrastive learning (GCL)~\cite{zhu2021graph,you2021graph} is one of the most widely used augmentation paradigms for recommendation~\cite{wu2021self,shuai2022review,cailightgcl}. The key insight behind GCL-based recommendation models is to obtain supervision signals from auxiliary learning tasks, which aim to supplement the main recommendation objective via SSL-enhanced co-training.

\begin{figure}[t]
    \centering
    \vspace{-0.1in}
    \includegraphics[width=0.95\columnwidth]{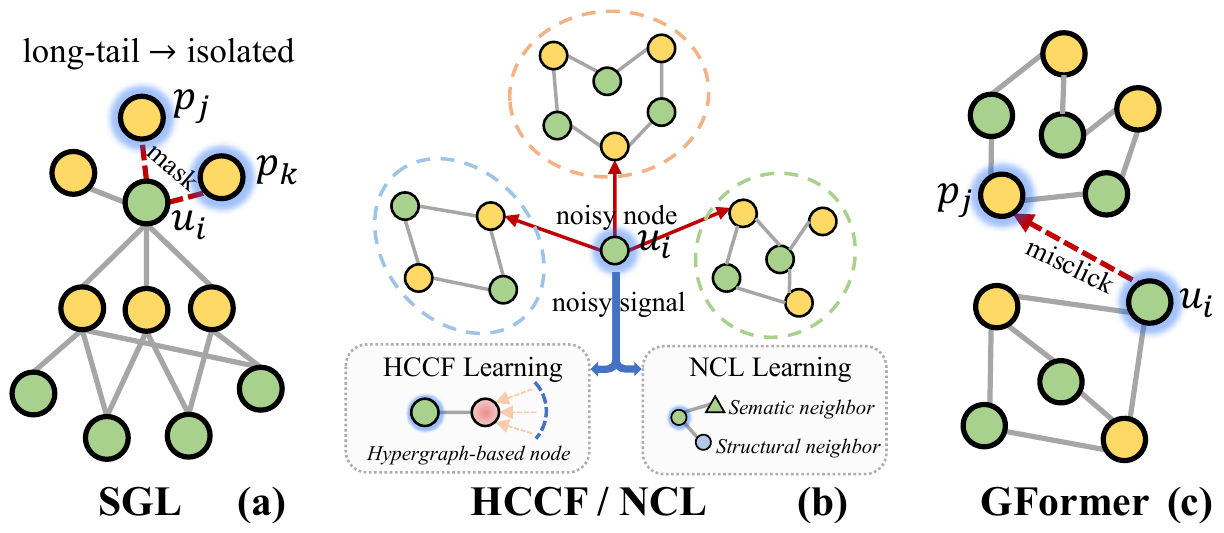}
    \vspace{-0.1in}
    \caption{Illustration for the graph augmentations generated by different self-supervised recommendation models.}
    \vspace{-0.25in}
    \label{fig:intro}
\end{figure}

Existing graph contrastive methods aim to maximize mutual information by achieving representation consistency between generated positive samples (\eg, user self-discrimination), and minimizing similarity between negative pairs (\eg, different users). Recent efforts have attempted to contrast different structural views of the user-item interaction graph with heuristic-based data augmentors, following the principle of mutual information maximization. For instance, SGL~\cite{you2021graph} proposes to corrupt graph structures by randomly removing user and item nodes as well as their connections to construct topological contrastive views. However, blindly corrupting graph topological structures can lead to the loss of crucial relations between users and items, such as unique user interaction patterns or limited labels of long-tail items (as illustrated in Figure~\ref{fig:intro}.(a)). Therefore, it is crucial to explicitly provide essential self-supervision signals for learning informative representations, which requires invariant rationales in the designed augmentors.


From the perspective of aligning local-level and global-level embeddings for augmentation, some research studies obtain semantic-related subgraph representations through various information aggregation techniques, such as hypergraph-based message passing in HCCF~\cite{xia2022hypergraph} and EM algorithm-based node clustering in NCL~\cite{lin2022improving}. However, due to their hand-crafted nature, the quality of augmentation is likely to be influenced by manually constructed hypergraph structures and user cluster settings. As a result, these augmentation schemes are insufficient to regularize the training process with useful self-supervised signals (\eg, truly negative pairs and hard augmented instances). Moreover, these manually designed contrastive methods can be easily misled by commonly existing noise (\eg, misclick behaviors~\cite{tian2022learning}, popularity bias~\cite{zhang2021causal}) (as shown in Figure~\ref{fig:intro}.(b)). Introducing augmented SSL information from biased data can amplify the noisy effects, which dilutes the learning of true user-item interaction patterns. Therefore, existing solutions may fall short in adapting the self-supervision process to changing practical recommendation environments.

Despite the advancements in SSL-enhanced recommender systems, a fundamental question remains poorly understood: \emph{What information is crucial and should be preserved for self-supervised augmentation in recommendation?} Motivated by the recent success of masked autoencoding (MAE) techniques in advancing self-supervised learning~\cite{he2022masked,chen2022sdae,hou2022graphmae}, this work explores the above question from the perspective of generative self-supervised augmentation with rationale-aware invariant representation learning. Unlike contrastive learning, the masked autoencoder paradigm directly adopts the reconstruction objective as the principled pretext task for data augmentation. It naturally avoids the limitations of manually-generated contrastive views for data augmentation discussed above. \\\vspace{-0.15in}


\noindent \textbf{Present Work}. In this work, we propose a new recommender system with \underline{G}raph Trans\underline{Former} to automatically distill masked self-supervised signals with invariant collaborative rationales. We take inspiration from rationale discovery~\cite{zhang2021improving,wudiscovering} to bridge the gap between the graph masked autoencoder with adaptive augmentation. Our \model\ makes full use of the power of Transformer in explicitly encoding pairwise relations to discover useful self-supervised signals benefiting the downstream recommendation task, with their own rationales explained. Specifically, we develop a topology-aware graph transformer to integrate into the user-item interaction modeling, enabling automated collaborative rationale discovery. In \model, the topological information of the user-item relation graph is treated as the global context in the form of graph positional encoding. To adapt \model\ to diverse recommendation environments, it learns to form appropriate interaction patterns as self-supervision signals, guided by task-adaptive collaborative rationale discovery. Our contributions can be summarized as follows:
\begin{itemize}[leftmargin=*]

\item This work revisits the self-supervised recommendation by exploring augmentation schemes from SSL-enhanced collaborative rationalization. We not only realize the automated data augmentors in SSL, but also provide rationale-aware understanding behind the self-supervised augmentation to improve model robustness.

\item We propose a principled approach for discovering invariant rationales with collaborative relations over the graph transformer. Task-aware adaptation is introduced to alleviate the issue of data-level variance. Then, the graph autoencoder is required to reconstruct the masked user-item interactions for augmentation.

\item We validate the effectiveness of our \model\ on several datasets. Compared with a variety of strong compared methods, our method consistently gains improvements across different settings. 


\end{itemize}

%% file: relate.tex
\section{Preliminaries and Related Work}

\noindent\textbf{Graph-based Collaborative Filtering}. Many recent studies have explored the use of graph representation learning in building graph-enhanced collaborative filtering (CF) models to capture high-order collaborative relations~\cite{he2020lightgcn,chen2020revisiting,xia2023graph}. In this scenario, we assume that there are $I$ users $\mathcal{U}=\{u_1, u_2,..., u_I\}$ and $J$ items $\mathcal{P}=\{p_1, p_2, ..., p_J\}$ in our recommendation system. The observed user behaviors are represented by an interaction matrix $\textbf{A}\in\mathbb{R}^{I\times J}$, where $a_{i,j}=1$ if an interaction between user $u_i$ and item $p_j$ is observed, and $a_{i,j}=0$ otherwise. To transform the interaction matrix into an interaction graph for graph-based CF, we define a graph $\mathcal{G}=\{\mathcal{V}, \mathcal{E}\}$, where $\mathcal{V}=\mathcal{U}\cup\mathcal{P}$ forms the node set, and $\mathcal{E}=\{e_{i,j}|a_{i,j}=1\}$ denotes the edge set corresponding to user-item interactions. Using these definitions, the graph-based CF can be abstracted as a prediction function over user-item interactions: $\hat{y}_{i,j} = f(\mathcal{G}; \mathbf{\Theta})$, where $\hat{y}_{i,j}$ is the predicted score for the unknown interaction between user $u_i$ and item $p_j$, and $\mathbf{\Theta}$ represents the model parameters. \\\vspace{-0.12in}


\noindent\textbf{GNN-enhanced CF Models}.
Graph neural networks (GNNs)~\cite{wu2019simplifying,kipfsemi} have become effective components for modeling user-item relationships in recommender systems. Typically, a GNN encoder is applied to generate user/item embeddings based on recursive message passing operations on the generated graph structures from user-item interactions~\cite{yang2021enhanced,wang2022profiling}. Earlier efforts adopt graph convolutional networks to map the interaction graph into latent embeddings, such as NGCF~\cite{wang2019neural} and Star-GCN~\cite{zhang2019star}. To simplify the graph message passing algorithm, recent approaches such as LightGCN~\cite{he2020lightgcn} and GCCF~\cite{chen2020revisiting} propose removing the non-linear transformation and activation for embedding propagation. Additionally, inspired by disentangled representation learning, latent intent disentanglement has been used to enhance graph neural networks for fine-grained user preference modeling, as demonstrated in DGCF~\cite{wang2020disentangled} and DisenHAN~\cite{wang2020disenhan}. Hyperbolic representation space has been introduced to improve graph collaborative filtering for user embedding, as shown in HGCF~\cite{sun2021hgcf}. Recent studies have also aimed to capture interaction heterogeneity for graph collaborative filtering, with approaches such as MBGCN~\cite{jin2020multi} and MBGMN~\cite{xia2021graph} designed to enable multiplex GNNs for learning multi-behavior interaction patterns. \\\vspace{-0.12in}


\noindent\textbf{Self-Supervised Learning for Recommendation}.
Recently, data augmentation with self-supervised learning (SSL) has emerged as a promising approach for mitigating the label scarcity and noise issue in recommender systems~\cite{yao2021self,yang2023debiased}. One important SSL paradigm is contrastive learning-based augmentation, where semantic-relevant instances are aligned with sampled positive pairs, while unrelated samples as negative pairs are pushed away. For example, random corruptions are performed on graph structures in SGL~\cite{wu2021self} and node embeddings in SLRec~\cite{yao2021self}. In addition, pre-defined embedding alignment methods are used to create views for embedding contrasting with heuristics, such as hypergraph construction in HCCF~\cite{xia2022hypergraph} and user clustering in NCL~\cite{lin2022improving}.


%% file: solution.tex
\section{Methodology}
\label{sec:solution}

\begin{figure*}
    \centering
    \includegraphics[width=1.0\textwidth]{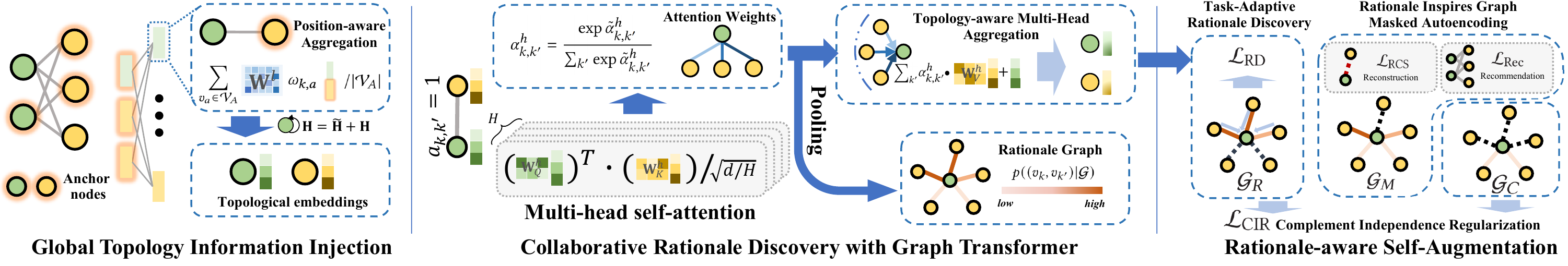}
    \vspace{-0.25in}
    \caption{Overall framework illustration of the proposed \model\ model. i) The collaborative rationale discovery is built upon the topology-aware graph transformer for interaction rationalization. ii) Position-aware message passing is enabled to encode pairwise user-item dependency with the global graph context enhancement. iii) Graph autoencoder aims at reconstructing the discovered collaborative rationales, with informative user-item interaction patterns for augmentation. iv) Task-adaptive self-supervision is realized with the awareness of main optimized objective derived from the target recommendation task.}
    \vspace{-0.1in}
    \label{fig:framework}
\end{figure*}


\subsection{Graph Invariant Rationale Learning}
\label{sec:graph_invariant}
To eliminate the impact of noisy features and enhance model interpretability, representation learning with rationalization has been explored to identify a subset of important features (\eg, language words~\cite{yu2019rethinking}, image pixels~\cite{zellers2019recognition}) that guide the model prediction results. Recently, rationalization learning techniques have been introduced into graph representation learning by discovering invariant rationales for important graph structural information to benefit downstream graph mining tasks~\cite{li2022learning,wudiscovering,li2022let}. In our graph-based CF scenario, our invariant rationale discovery scheme is designed to find a subset of graph structures that best guide the self-supervised augmentation for the downstream recommendation task with rationalization. Our invariant rationale discovery with graph collaborative relationships aims to optimize the following objective from two viewpoints: \emph{performance sufficiency} and \emph{complement independence}. This objective is formally given as: 
\begin{align}
    \min D\left(f(R(\mathcal{G})), f(\mathcal{G})\right) + I\left(R(\mathcal{G}), C(R(\mathcal{G}))\right)
\end{align}
\noindent $f(\cdot)$ denotes the predictive function, while $R(\cdot)$ and $C(\cdot)$ denote the rationale and complement of the rationale for the input graph $\mathcal{G}$, respectively. Specifically, for achieving performance sufficiency, the first term aims to minimize the performance difference between using the rationale $R(\mathcal{G})$ and the entire graph $\mathcal{G}$. Here, $D(\cdot)$ represents the difference measurement function. By doing so, important structural information of graph collaborative relations is well-preserved in our learned rationale $R(\mathcal{G})$. Additionally, in pursuit of complement independence by mitigating noisy signals, the second term seeks to minimize the dependency of the complement graph structures $C(R(\mathcal{G}))$ and rationale $R(\mathcal{G})$. With this objective, the complement of our discovered rationales has little influence on the label prediction. Hence, our graph rationale discovery can exploit the invariant relationships between users and items while alleviating the noisy effects of spurious interactions. 


\subsubsection{\bf Graph Collaborative Rationale Discovery}
To enable noise-resistant self-supervision for augmentations, our \model\ aims at automatically distilling important graph structures over the interaction graph $\mathcal{G}$, \ie, the collaborative rationales. To generate the informative interaction subgraph structure, our collaborative rationale discovery is designed to estimate the following probability of subgraph $\mathcal{G}_R$ being the rationale of interaction graph $\mathcal{G}$:
\begin{align}
    &p(R(\mathcal{G})=\mathcal{G}_R)=\prod_{e\in \mathcal{E}_R} p(e|\mathcal{G})\prod_{e_{'}\in\mathcal{E}_C}\left(1- p(e_{k'}|\mathcal{G})\right)\nonumber\\
    &\mathcal{G}_R=\{\mathcal{V}, \mathcal{E}_R\},~~~ \mathcal{G}_R\sim p(R(\mathcal{G})=\mathcal{G}_R),~~~|\mathcal{E}_R|=\rho_R\cdot|\mathcal{E}| \nonumber\\
    &\mathcal{G}_C=\{\mathcal{V}, \mathcal{E}_C\},~~~ \mathcal{E}_C = \{e_{'}|e_{'}\in\mathcal{E}, e_{'}\notin\mathcal{E}_R\}
\end{align}
\noindent where $\rho_R$ denotes the proportion of interaction edges selected as the collaborative rationales $\mathcal{G}_R$, and $\mathcal{G}_C$ is defined as the subgraph containing the edges that are not part of $\mathcal{G}_R$. Here, we define $e$ and $e_{'}$ to denote user-item interactions in the rationale and complement subgraphs, respectively. To estimate the distribution probability described above, our \model\ proposes to infer the probability of individual edge $p(e|\mathcal{G})$ and $p(e_{'}|\mathcal{G})$ being identified as part of the rationale. With a graph encoder for node embeddings, the parameterized rationale generator is formally generalized as follows:
\begin{align}
    p(e|\mathcal{G})\leftarrow \textbf{GT} \left(\mathcal{G}, \textbf{TE} (\textbf{H};\mathbf{\Theta}_\textbf{TE}); \mathbf{\Theta}_\textbf{GT}\right);~~
    \mathop{\arg\max}\limits_{\mathbf{\Theta}_\textbf{TE}, \mathbf{\Theta}_\textbf{GT}} \mathcal{L}_{\text{RD}}
\end{align}
\noindent Inspired by the design of dependency rationalization of self-attention in Transformer, our graph encoder $\textbf{GT}(\cdot)$ is built upon a Graph Transformer architecture, which will be elaborated on in Section~\ref{sec:graph_transformer}. To inject the global topological context into the invariant rationale discovery process, we design the graph topology embedding module $\textbf{TE}(\cdot)$ to capture the collaborative effects across the entire graph, as detailed in Section~\ref{sec:topology}. Specifically, $\textbf{H}\in\mathbb{R}^{(I+J)\times d}$ represents the embedding table containing the representations of all $I$ user nodes and $J$ item nodes. The learnable parameters of the embedding functions $\textbf{GT}(\cdot)$ and $\textbf{TE}(\cdot)$ are denoted by $\mathbf{\Theta}_{\textbf{GT}}$ and $\mathbf{\Theta}_{\textbf{TE}}$, respectively. In our learning process, $\mathbf{\Theta}_{\textbf{GT}}$ and $\mathbf{\Theta}_{\textbf{TE}}$ are inferred by optimizing the BPR-based objective function $\mathcal{L}_{RD}$ for the collaborative filtering task. This enables task-adaptive rationale discovery for SSL augmentation in our \model\ model.

\subsubsection{\bf Global Topology Information Injection}
\label{sec:topology}
Motivated by the power of position-aware graph neural networks~\cite{you2019position} in capturing global relational information, our \model\ proposes to enhance collaborative rationale discovery by preserving high-order user/item dependencies. We begin by sampling a set of anchor nodes $\mathcal{V}_A\subset\mathcal{V}$ from the user-item interaction graph $\mathcal{G}=\{\mathcal{V}, \mathcal{E}\}$. To represent the global topological embeddings of users and items based on their connectivities with anchor nodes, we calculate the distance $d_{k,a}$ between the target node $v_k$ and each anchor node $v_a$, where distance is defined as the minimum number of edges that must be traversed to go from $v_k$ to $v_a$ in $\mathcal{G}$. Given the calculated distances, we derive the correlation weight $\omega_{k,a}$ for each pair of target-anchor nodes $(k,a)$ as follows:
\begin{align}
    \omega_{k,a}
    = \left\{\begin{matrix} 
    \frac{1}{d_{k,a}+1} \quad if \quad d_{k,a}\leqslant q \\ 
    0 \quad\,\qquad otherwise
    \end{matrix}\right.
\end{align}
\noindent $q$ represents the maximum value allowed for the correlation weight between any target and anchor node, which is used for normalization purposes. The node correlation weights are then normalized to the range of $[0, 1]$. With the weight $\omega_{k,a}$, we refine the target node embedding by considering the correlation weights between the target node $k$ and each anchor node $v_a\in\mathcal{V}_A$:
\begin{align}
    \tilde{\textbf{h}}_k^{l} = \sum_{v_a\in\mathcal{V}_A} \textbf{W}^{l} \cdot \omega_{k,a}\cdot [\tilde{\textbf{h}}_k^{l-1}||\tilde{\textbf{h}}_a^{l-1}] / |\mathcal{V}_A|
\end{align}
\noindent Here, $\tilde{\textbf{h}}_k^{l}, \tilde{\textbf{h}}_k^{l-1}\in\mathbb{R}^d$ denote the embeddings of node $v_k$ in the $l$-th and $(l-1)$-th graph propagation layer, respectively. To represent the global topological context based on anchor nodes, we define $\textbf{H}^{l_T}$ to denote the global topological embedding matrix in the $l$-th layer. $\textbf{W}^{l}\in\mathbb{R}^{d\times 2d}$ is a learnable transformation matrix. $[\cdot||\cdot]$ denotes vector concatenation. After $L$ graph information propagation steps, the embeddings in $\tilde{\textbf{H}}^{L}$ preserve the high-order topological information. We then inject this information into the id-corresponding embeddings to obtain the topological embeddings, as follows:
\begin{align}
     \bar{\textbf{H}} = \textbf{TE}(\textbf{H};\{\textbf{W}_T^{l_T}\})
\end{align}
\noindent In this way, our parameterized rationale generator can capture global collaborative relationships and identify informative patterns of interactions between users and items for SSL augmentation.

\subsubsection{\bf Rationale Discovery with Graph Transformer}
\label{sec:graph_transformer}

Our rationale discovery aims to extract informative patterns of user-item interactions that can be used for self-supervised augmentation in changing recommendation environments with limited supervision labels. To address this challenge, we draw inspiration from the Transformer architecture and its core self-attention mechanisms. Specifically, we propose a novel approach to learning environment-invariant user preference information as generative self-supervision signals with selective augmentation. This design allows our \model\ to mitigate the noise induced by observational behavior data, which is prone to contain biases and confounding factors that can negatively affect recommendation performance.

Our parameterized rationale discovery module is built over graph transformer framework to encode implicit label-invariant node-wise relations as selected rationales. To incorporate the positional information of user and item nodes into the topology learning process, we feed the global topology-aware node embeddings $\bar{\textbf{H}}$ into the multi-head self-attention mechanism for rationalization. Specifically, we learn the correlation between node $v_k$ and $v_{k'}$ with respect to the $h$-th attention head as follows:
\begin{align}
    \alpha_{k,k'}^h = \frac{\exp\tilde{\alpha}^h_{k,k'}}{\sum_{k'}\exp\tilde{\alpha}^h_{k,k'}};~~~
    \tilde{\alpha}^h_{k,k'} = \frac{(\textbf{W}_Q^h\cdot \bar{\textbf{h}}_k)^\top\cdot(\textbf{W}_K^h\cdot\bar{\textbf{h}}_{k'})}{\sqrt{d/H}}
\end{align}
\noindent Here, $\textbf{W}_Q^h$ and $\textbf{W}_K^h\in \mathbb{R}^{\frac{d}{H}\times d}$ represent the transformations used to obtain the query and key embeddings for calculating the attention scores. Since the attention scores encoded by our graph transformer capture the strength of node-wise dependencies, we aggregate the multi-head scores to obtain the probability scores, $p((v_k,v_{k'}) | \mathcal{G})$, of graph edges, such as $v_k$--$v_{k'}$, being selected as rationales. These rationales correspond to the subset of important user-item interaction patterns that best illuminate the user preference learning process with invariant representations, which is presented as:
\begin{align}
    p((v_k,v_{k'})|\mathcal{G}) = \frac{\bar{\alpha}_{k,k'}}{\sum_{(v_k,v_{k'})\in\mathcal{E}} \bar{\alpha}_{k,k'}};~~~
    \bar{\alpha}_{k,k'} = \sum_{h=1}^H \alpha_{k,k'}^h / H
\end{align}
To sample a rationale estimated by our topology-aware graph transformer, we individually sample $\rho_R\cdot|\mathcal{E}|$ edges from the edge set $\mathcal{E}$ according to their probability scores, $p((v_k, v_{k'})|\mathcal{G})$. Here, the hyperparameter $\rho_R\in\mathbb{R}$ controls the size of the subset of important edges that are selected for rationalization.

\subsubsection{\bf Task-Adaptive Rationale Discovery}
To perform task-level adaptation in our rationale discovery, our \model\ is a task-adaptive rationale discovery paradigm that can perform task-specific rationalization to provide customized recommendations. Specifically, our model leverages the embeddings and the distilled rationales from the graph transformer to generate predictions for users' preferences over items. This process is formally given as follows:
\begin{align}
    \bar{y}_{i,j} &= \textbf{z}_i^{L\top} \cdot\textbf{z}_j^{L};~~~~\textbf{z}_k^{L} = \sum_{(v_k, v_{k'})\in\mathcal{E}_R} \beta_{k,k'}\cdot\textbf{z}_{k'}^{l-1};\nonumber\\
    {\textbf{Z}}^0 &= \textbf{GT}(\mathcal{G}, \textbf{TE}(\textbf{H}))= \mathop{\Big|\Big|}\nolimits_{h=1}^H \sum\nolimits_{k'} \alpha_{k,k'}^h \textbf{W}_V^h\bar{\textbf{h}}_{k'} + \bar{\textbf{h}}_k
\end{align}
\noindent $\bar{y}_{i,j}\in\mathbb{R}$ denotes the predicted probability of user $u_i$ adopting item $p_j$. The embeddings $\textbf{z}_i^{L}$ and $\textbf{z}_j^{}\in\mathbb{R}^d$ are used to make predictions on user-item interactions, while $\textbf{z}_k^{L}$ for a vertex $v_k$ is obtained through an $L$-layer LightGCN~\cite{he2020lightgcn}. Here $\mathcal{E}_R$ denotes the edge set of the sampled rationale graph $\mathcal{G}_R$. $\beta_{k,k'}=1 / \sqrt{d_k d_{k'}}$ denotes the Lapalacian normalization with degrees $d_k$ and $d_{k'}$ of node $v_k$ and $v_{k'}$. The 0-order embeddings $\textbf{Z}^0$ are obtained through multi-head embedding aggregation in the topology-aware graph transformer, where $H$ is the number of attention heads, and $\textbf{W}_V^h\in\mathbb{R}^{d/H\times d}$ denotes the value transformation in the self-attention. We employ a residual connection to also use the topology-aware embeddings $\bar{\textbf{h}}_{k'}$ as input. With the predicted probability score for each $(u_i, y_j)$ interaction, we apply the following BPR loss to guide the rationale discovery process and optimize the objective of the downstream task:
\begin{align}
    \label{eq:l_rd}
    \mathcal{L}_{\text{RD}} = \sum_{(u_i, p_j^+, p_j^-)} -\log\text{sigm}(\bar{y}_{i,j^+} - \bar{y}_{i, j^-})
\end{align}
\noindent A pair-wise training triplet is formed by sampling a user and items such that $u_i\in\mathcal{U}$ and $v_j^+,v_j^-\in\mathcal{P}$, and the triplet satisfies $ (u_i, p_j^+)\in\mathcal{E}$ and $(u_i, p_j^-)\notin\mathcal{E}$. The sigmoid function is represented by $\text{sigm}(\cdot)$. To incorporate task-specific knowledge, a topology-aware graph transformer is employed. This transformer provides task-aware parameter learning to customize the collaborative rationale discovery for different recommendation scenarios. As a result, the collaborative rationale discovery satisfies the sufficiency principle~\cite{yu2019rethinking} with the objective of $f(R(\mathcal{G}))=f(\mathcal{G})$.


\subsection{Rationale-Aware Self-Augmentation}
\subsubsection{\bf Rationales Inspire Graph Masked Autoencoding} Our proposed self-distillation paradigm for discovering collaborative rationales involves performing self-augmentation over the distilled informative user-item interaction patterns through graph masked autoencoding. To achieve this, we configure our \model\ with the rationale-aware mask autoencoder, which masks identified rationables from the interaction graph for autoencoding-based reconstruction. To sample the masked graph structure $\mathcal{G}_M = \{\mathcal{V}, \mathcal{E}_M\}$, we use the reciprocal of the rationale scores. This allows us to mask the most important rationale structures, as shown below:
\begin{align}
    &\mathcal{E}_M\sim p_M(\mathcal{E}_M|\mathcal{G})= \prod_{(v_k, v_{k'})\in\mathcal{E}_M} \alpha^M_{k,k'}\prod_{(v_k,v_{k'})\in \mathcal{E}\backslash\mathcal{E}_M}\alpha_{k,k'}^M\nonumber\\
    &|\mathcal{E}_M|=\rho_M |\mathcal{E}|;~~ \alpha_{k,k'}^M = \frac{\bar{\alpha}_{k,k'}^M}{\sum\limits_{(v_k,v_{k'})\in\mathcal{E}} \bar{\alpha}_{k,k'}^M};~~ \bar{\alpha}_{k,k'}^M=\frac{1}{\bar{\alpha}_{k,k'}+\epsilon}
\end{align}
\noindent $p_M(\cdot)$ is the probability of sampling edges in the masked graph. $\alpha_{k,k'}^M$ is the probability of selecting an edge between nodes $v_k$ and $v_{k'}$ in the mask generator. $\bar{\alpha}_{k,k'}^M$ is the un-normalized attention score calculated using the reciprocal of the weights $\alpha_{k,k'}$. A small value $\epsilon$ is added to avoid a zero denominator. The masked graph has a higher edge density than the rationale graph to only remove the most important rationale edges for noise-resistant autoencoding. The masked graph $\mathcal{G}_M$ with edge set $\mathcal{E}_M$ is then used as input for the autoencoder network, which is presented as follows:
\begin{align}
    \textbf{S} = \textbf{GT}(\mathcal{G}_M, \textbf{TE}(\bar{\textbf{S}}^{L})); ~~~
    \bar{\textbf{s}}^{l} = \sum_{(v_k,v_{k'})\in\mathcal{E}_M} \beta_{k,k'} \cdot \bar{\textbf{s}}_{k'}^{l_1}
\end{align}
\noindent $\textbf{S}\in\mathbb{R}^{(I+J)\times d}$ represents the final embeddings in the autoencoder. $\textbf{GT}(\cdot)$ and $\textbf{TE}(\cdot)$ denote our graph transformer network and the topological information encoder, respectively. We enhance the initial embeddings with $L$-order local node embeddings $\bar{\textbf{S}}^{L}$, encoded from LightGCN. $\bar{\textbf{S}}^0$ is initialized with the id-corresponding embeddings $\textbf{H}$. The embeddings $\textbf{S}$ are used for training the reconstruction of the masked user-item interactions. This can be expressed as:
\begin{align}
    \mathcal{L}_\text{MAE} = \sum_{(v_k, v_{k'})\in\mathcal{E}\backslash\mathcal{E}_M} -\tilde{y}_{k,k'};~~~~~
    \tilde{y}_{k,k'}=\textbf{s}_{k}^\top \textbf{s}_{k'}
\end{align}
\noindent $\mathcal{L}_\text{MAE}$ is the training objective for reconstructing the masked interaction patterns. $\tilde{y}_{k,k'}$ represents the predicted scores for edge $(v_k, v_{k'})$ on graph $\mathcal{G}$. Inspired by our collaborative rationale discovery, our graph masked autoencoder is trained to reconstruct important interaction patterns that are adaptable to downstream recommendation tasks. Our rationale-aware augmentation approach prevents our generative SSL from being influenced by noisy edges.

\subsubsection{\bf Complement Independence Modeling}
To achieve complement independence in rationale discovery (as discussed in Section~\ref{sec:graph_invariant}), we introduce a learning component to encourage independence between the distilled collaborative rationales and their corresponding complements, thereby reducing information redundancy. This is done through contrastive regularization, where we minimize the mutual information between the rationale graph $\mathcal{G}_R$ and a sampled complement graph $\mathcal{G}_C$. The complement graph is sampled in a manner similar to graph masking, but with a different sampling rate $\rho_C << \rho_M$ to identify noisy edges. The complement graph $\mathcal{G}_C=\{\mathcal{V}, \mathcal{E}_C\}$ is generated as follows:
\begin{align}
    \mathcal{E}_C\sim p_C(\mathcal{E}_C|\mathcal{G})=p_M(\mathcal{E}_C|\mathcal{G});~~~
    |\mathcal{E}_C| = \rho_C \cdot |\mathcal{E}|
\end{align}
\noindent To ensure that the complement graph $\mathcal{G}_C$ does not contain non-noise edges that could affect the independence regularization, we use a low sampling rate $\rho_C$. We then apply the following loss to minimize the similarities between the rationale graph $\mathcal{G}_R$ and the complement graph $\mathcal{G}_C$ in high-order representations:
\begin{align}
    &\mathcal{L}_\text{CIR}= \log\sum_{v_k\in\mathcal{V}} \exp\cos(\textbf{e}_k^R, \textbf{e}_{k}^C) / \tau \nonumber\\
    \textbf{E}^R = &\textbf{L-GCN}^{L}(\textbf{H}, \mathcal{G}_R);~~~
    \textbf{E}^C = \textbf{L-GCN}^{L}(\textbf{H}, \mathcal{G}_C)
\end{align}
\noindent $\textbf{L-GCN}^{L}(\cdot)$ represents the stacking of $L$ graph layers in LightGCN~\cite{he2020lightgcn} for recursively passing messages over the input graph ($\mathcal{G}_R$ and $\mathcal{G}_C$). $\tau$ is the hyperparameter for the temperature coefficient. The contrastive independence regularization $\mathcal{L}_\text{CIR}$ pushes the embeddings of the rationale embeddings $\textbf{e}_k^R$ and the complement embeddings $\textbf{e}_k^C$ away from each other for all nodes. This enhances the model's ability to encourage independence between the discovered rationales and complements, thereby improving the noise mitigation ability in our \model\ for SSL-based augmentation.

\subsubsection{\bf SSL-Augmented Model Optimization}
During the training phase, we use the embeddings $\textbf{S}\in\mathbb{R}^{(I+J)\times d}$ to make predictions for training the recommender. The following point-wise loss function is minimized for model training:
\begin{align}
    \mathcal{L}_\text{Rec} = \sum_{a_{i,j}=1} -\log\frac{\exp\textbf{s}_i^\top \textbf{s}_j}{\sum_{p_{j'}\in\mathcal{P}}\exp \textbf{s}_i^\top \textbf{s}_{j'}}
\end{align}
\model\ maximizes the predictions for all positive user-item interactions and minimizes the predictions for all negative user-item interactions as a contrast. During the testing phase, we replace the masked graph $\mathcal{G}_M$ in \model\ with the observed interaction graph $\mathcal{G}$ and predict the relations between user $u_i$ and item $p_j$ by $\hat{y}_{i,j}=\textbf{s}_i^\top\textbf{s}_j$. By combining the multiple training objectives, our \model\ is optimized to minimize the following overall objective:
\begin{align}
    \mathcal{L}=\mathcal{L}_\text{Rec} + \mathcal{L}_\text{RCS} + \lambda_1 \cdot \mathcal{L}_\text{RD} + \lambda_2 \cdot \mathcal{L}_\text{CIR} + \lambda_3 \cdot \|\mathbf{\Theta}\|_\text{F}^2
\end{align}
\noindent $\lambda_1, \lambda_2,$ and $\lambda_3$ are hyperparameters used for loss balancing. The last term is the Frobenius-norm regularization for the parameters.

\subsection{Discussion of Time and Space Complexity}
\noindent \textbf{Time Complexity}. Our \model\ employs graph transformer for collaborative rationale discovery and LightGCN as our graph encoder for rationale subgraph structures. The former takes $\mathcal{O}(L_T\times (I+J)\times d^2)$ complexity for embedding transformation and $\mathcal{O}(L_T\times (\rho_R + \rho_M)\times |\mathcal{E}| \times d)$ for information propagation and aggregation. LightGCN requires $\mathcal{O}((L_C\rho_R + L_A\rho_M + L_I\rho_R)\times |\mathcal{E}|\times d)$ cost. \\\vspace{-0.12in}



\noindent \textbf{Space Complexity}. Our collaborative rationale discovery module is built directly upon the graph encoder--Graph Transformer, which means that no additional rationale learning parameters are needed compared to other rationale graph structure learning methods (\eg~\cite{li2022let, luo2021learning}). As a result, our \model\ model requires a smaller space cost (i.e., $\mathcal{O}((I+J)\times d + d^2)$) than these methods.



%% file: eval.tex
\section{Evaluation}
\label{sec:eval}

\begin{table}[t]
    \centering
    \small
    \caption{Statistics of the experimental datasets.}
    \vspace{-0.12in}
    \begin{tabular}{ccccc}
        \toprule
        Dataset & \#Users & \#Items & \#Interactions & Density\\
        \midrule
        Yelp & 42,712 & 26,822 & 182,357 & 1.6$e^{-4}$\\
        Ifashion & 31,668 & 38,048 & 618,629 & 5.1$e^{-4}$\\
        LastFM & 1,889 & 15,376 & 51,987 & 1.8$e^{-3}$\\
        \hline
    \end{tabular}
    \vspace{-0.17in}
    \label{tab:data statistics}
\end{table}

\begin{table*}[t]
\centering
\small
\setlength{\tabcolsep}{1mm}
\caption {Performance comparison between our proposed \model\ and all baselines on Ifashion, Yelp, LastFM datasets.}
\vspace{-0.1in}
\begin{tabular}{ccccclcccclcccccc}
\hline
Datasets                  & Metric    & BiasMF & NCF    & AutoRec & PinSage & NGCF   & GCCF   & LightGCN & EGLN   & SLRec  & NCL  & HCCF   & SGL    & \model\   & p-val.  \\ \hline
\multirow{6}{*}{Yelp}     & Recall@10 & 0.0122 & 0.0166 & 0.0230  & 0.0278  & 0.0438 & 0.0484 & 0.0422   & 0.0458 & 0.0418 & 0.0493 & 0.0518 & 0.0522 & \textbf{0.0562} & 2.8e-6  \\ \cline{2-16} 
                          & NDCG@10   & 0.0070 & 0.0101 & 0.0133  & 0.0171  & 0.0269 & 0.0296 & 0.0254   & 0.0278 & 0.0258 & 0.0301 & 0.0318 & 0.0319 & \textbf{0.0350} & 9.8e-9  \\ \cline{2-16} 
                          & Recall@20 & 0.0198 & 0.0292 & 0.0410  & 0.0454  & 0.0678 & 0.0754 & 0.0761   & 0.0726 & 0.0650 & 0.0806 & 0.0789 & 0.0815 & \textbf{0.0878} & 5.2e-8 \\ \cline{2-16} 
                          & NDCG@20   & 0.0090 & 0.0138 & 0.0186  & 0.0224  & 0.0340 & 0.0378 & 0.0373   & 0.0360 & 0.0327 & 0.0402 & 0.0391 & 0.0410 & \textbf{0.0442} & 1.6e-6 \\ \cline{2-16} 
                          & Recall@40 & 0.0303 & 0.0442 & 0.0678  & 0.0712  & 0.1047 & 0.1163 & 0.1031   & 0.1121 & 0.1026 & 0.1192 & 0.1244 & 0.1249 & \textbf{0.1328} & 8.3e-11 \\ \cline{2-16} 
                          & NDCG@40   & 0.0117 & 0.0167 & 0.0253  & 0.0287  & 0.0430 & 0.0475 & 0.0413   & 0.0456 & 0.0418 & 0.0485 & 0.0510 & 0.0517 & \textbf{0.0551} & 7.8e-9  \\ \hline
\multirow{6}{*}{Ifashion} & Recall@10 & 0.0302 & 0.0268 & 0.0309  & 0.0291  & 0.0375 & 0.0373 & 0.0437   & 0.0473 & 0.0373 & 0.0474 & 0.0489 & 0.0512 & \textbf{0.0542} & 6.3e-7  \\ \cline{2-16} 
                          & NDCG@10   & 0.0281 & 0.0253 & 0.0264  & 0.0276  & 0.0350 & 0.0352 & 0.0416   & 0.0438 & 0.0353 & 0.0446 & 0.0464 & 0.0487 & \textbf{0.0520} & 2.0e-6  \\ \cline{2-16} 
                          & Recall@20 & 0.0523 & 0.0451 & 0.0537  & 0.0505  & 0.0636 & 0.0639 & 0.0751   & 0.0787 & 0.0633 & 0.0797 & 0.0815 & 0.0845 & \textbf{0.0894} & 1.5e-7  \\ \cline{2-16} 
                          & NDCG@20   & 0.0360 & 0.0306 & 0.0351  & 0.0352  & 0.0442 & 0.0445 & 0.0528   & 0.0549 & 0.0444 & 0.0558 & 0.0578 & 0.0603 & \textbf{0.0635} & 6.3e-5  \\ \cline{2-16} 
                          & Recall@40 & 0.0858 & 0.0785 & 0.0921  & 0.0851  & 0.1062 & 0.1047 & 0.1207   & 0.1277 & 0.1043 & 0.1283 & 0.1306 & 0.1354 & \textbf{0.1424} & 9.0e-9 \\ \cline{2-16} 
                          & NDCG@40   & 0.0474 & 0.0423 & 0.0483  & 0.0470  & 0.0585 & 0.0584 & 0.0677   & 0.0715 & 0.0582 & 0.0723 & 0.0744 & 0.0773 & \textbf{0.0818} & 9.9e-8 \\ \hline
\multirow{6}{*}{LastFM}   & Recall@10 & 0.0609 & 0.0574 & 0.0543  & 0.0899  & 0.1257 & 0.1230 & 0.1490   & 0.1133 & 0.1175 & 0.1491 & 0.1502 & 0.1496 & \textbf{0.1573} & 5.8e-7  \\ \cline{2-16} 
                          & NDCG@10   & 0.0696 & 0.0645 & 0.0599  & 0.1046  & 0.1489 & 0.1452 & 0.1739   & 0.1263 & 0.1384 & 0.1745 & 0.1773 & 0.1775 & \textbf{0.1831} & 1.8e-6  \\ \cline{2-16} 
                          & Recall@20 & 0.0980 & 0.0956 & 0.0887  & 0.1343  & 0.1918 & 0.1816 & 0.2188   & 0.1823 & 0.1747 & 0.2196 & 0.2210 & 0.2236 & \textbf{0.2352} & 5.0e-8  \\ \cline{2-16} 
                          & NDCG@20   & 0.0860 & 0.0800 & 0.0769  & 0.1229  & 0.1759 & 0.1681 & 0.2018   & 0.1557 & 0.1613 & 0.2021 & 0.2047 & 0.2070 & \textbf{0.2145} & 1.7e-8  \\ \cline{2-16} 
                          & Recall@40 & 0.1450 & 0.1439 & 0.1550  & 0.1990  & 0.2794 & 0.2649 & 0.3156   & 0.2747 & 0.2533 & 0.3130 & 0.3184 & 0.3194 & \textbf{0.3300} & 4.3e-7  \\ \cline{2-16} 
                          & NDCG@40   & 0.1067 & 0.1055 & 0.1031  & 0.1515  & 0.2146 & 0.2049 & 0.2444   & 0.1966 & 0.1960 & 0.2437 & 0.2458 & 0.2498 & \textbf{0.2567} & 4.2e-7  \\ \hline
\end{tabular}
\label{tab:overall}
\end{table*}

In this section, we conduct extensive experiments for model evaluation to answer the following key research questions:
\begin{itemize}[leftmargin=*]

\item \textbf{RQ1}: How effective is our \model\ compared to various state-of-the-art (SOTA) recommendation models? \\\vspace{-0.12in}

\item \textbf{RQ2}: How does the model performance change if we substitute key modules of \model\ with different naive implementations?\\\vspace{-0.12in}
\item \textbf{RQ3}: How does our rationale-aware graph transformer perform against data noise and data scarcity issues?\\\vspace{-0.12in}
\item \textbf{RQ4}: What is the training efficiency of the proposed \model?\\\vspace{-0.12in}
\item \textbf{RQ5}: How do key parameters affect the model performance?\\\vspace{-0.12in}
\item \textbf{RQ6}: How does our collaborative rationale discovery paradigm realize the interpretability of user-item interaction patterns?

\end{itemize}

\subsection{Experimental Setup}

\subsubsection{\bf Datasets.} 
We conduct experiments on three widely-used real-world datasets for evaluating recommender systems: Yelp, Ifashion, and LastFM. The Yelp dataset is used for recommending businesses venues to users, and it is collected from the well-known Yelp platform. Ifashion is a fashion outfit dataset collected by Alibaba, while LastFM is a dataset that tracks user interaction activities in music applications and internet radio sites. Table \ref{tab:data statistics} summarizes the statistics of the three experimental datasets.

\subsubsection{\bf Evaluation Protocols.} 
We split the observed interactions of each dataset into training set, validation set, and test set using a ratio of \(0.70:0.05:0.25\). To measure the recommendation accuracy for each user over the whole item set, we adopted the all-rank protocol, following~\cite{wu2021self,xia2022hypergraph}. This protocol helps alleviate the evaluation bias introduced by negative sampling. We evaluate all models using two representative metrics: Recall Ratio (\emph{Recall@K}) and Normalized Discounted Cumulative Gain (\emph{NDCG@K}), with \(K = 10, 20, 40\).

\subsubsection{\bf Baseline Methods.} To comprehensively study the performance of \model, we compare it with many baseline methods covering various techniques for collaborative filtering.

\vspace{0.1in}
\noindent\textbf{Non-GNN Collaborative Filtering Approaches}. We first include several conventional CF methods as benchmarks for comparison.
\begin{itemize}[leftmargin=*]
\item \textbf{BiasMF}~\cite{koren2009matrix}: This is a method based on matrix factorization which maps users and items to vector representations in the latent space and takes their bias score into consideration. \\\vspace{-0.12in}

\item \textbf{NCF}~\cite{he2017neural}: This method uses neural networks with multiple layers to encode non-linear features of user-item interactions.\\\vspace{-0.12in}

\item \textbf{AutoRec}~\cite{sedhain2015autorec}: This model adopts the autoencoder structure and learns embeddings through reconstructing observed interactions.
\end{itemize}
\noindent\textbf{GNN-based Recommendation Methods without SSL}. Graph neural networks (GNNs) have shown their effectiveness in injecting high-order collaborative signals into user and item embeddings. For this research line, we compared our \model\ with representative GNN-enhanced recommendation models that use various message passing schemes for performance evaluation.
\begin{itemize}[leftmargin=*]
\item \textbf{PinSage}~\cite{ying2018graph}: This model leverages graph convolutional networks with random-walk-based message passing to encode the user-item interaction graph with high efficiency.\\\vspace{-0.12in}

\item \textbf{NGCF}~\cite{wang2019neural}: This model captures high-order collaborative information through multiple layers of graph neural networks.\\\vspace{-0.12in}

\item \textbf{LightGCN}~\cite{he2020lightgcn}: This approach simplifies the architecture of NGCF and employs a light-weighted convolutional graph encoder for better representation learning and model training.\\\vspace{-0.12in}

\item \textbf{GCCF}~\cite{chen2020revisiting}: This model also introduces several improvements to GCN-based CF methods, including omitting the non-linear transformation and applying residual connections.\\\vspace{-0.12in}
\end{itemize}
\noindent\textbf{SSL-enhanced Recommendation Models}. For comprehensive model evaluation, we included many recent SSL-enhanced recommender systems as baselines. In these models, different augmentation strategies are designed to provide self-supervision signals.
\begin{itemize}[leftmargin=*]
\item \textbf{EGLN}~\cite{yang2021enhanced}: This model incorporates a node embedding learning module and a graph structure learning module, and encourages them to learn from each other for better representations.\\\vspace{-0.12in}

\item \textbf{SLRec}~\cite{yao2021self}: This approach conducts contrastive learning between node features to regularize the recommendation learning.\\\vspace{-0.12in}

\item \textbf{NCL}~\cite{lin2022improving}: This model first uses an EM algorithm to perform clustering over users, and then conducts neighborhood-enriched contrastive learning within each cluster.\\\vspace{-0.12in}

\item \textbf{SGL}~\cite{wu2021self}: This model uses random data augmentation operators (\eg, edge dropping, node dropping, and random walks) to construct views over interaction structures for contrastive learning.\\\vspace{-0.12in}

\item \textbf{HCCF}~\cite{xia2022hypergraph}: This is a state-of-the-art model that conducts contrastive learning through constructing hypergraph-based global and local views for modeling global relations.\\\vspace{-0.12in}

\end{itemize}

\subsubsection{\bf Hyperparameter Settings} We implement our \model\ using PyTorch. We use the Adam optimizer for parameter learning with a learning rate of \(1e^{-3}\) and no learning rate decay. For model hyperparameters of \model, we set the embedding size to \(d=32\) by default, the size of the anchor node set to \(|\mathcal{V}_A|=32\), and tuned the graph rationale keep rate \(\rho_R\) in \(\{0.5, 0.6, 0.7, 0.8, 0.9\}\). For coefficients of different loss terms, we search for \(\lambda_1\) in \(\{0.5, 1, 2, 4, 8\}\), \(\lambda_2\) in the range \(\{1, 1e^{-1}, 1e^{-2}, 1e^{-3}, 1e^{-4}\}\), and \(\lambda_3\) in the range \(\{1e^{-3},1e^{-4},1e^{-5},1e^{-6},1e^{-7},1e^{-8}\}\) respectively. We chose the number of graph transformer layers in the range of \(\{1,2,3,4,5,6\}\), and the number of graph convolutional layers in \(\{1,2,3,4,5\}\).

\begin{figure}[t]
    \centering
    \includegraphics[width=\columnwidth]{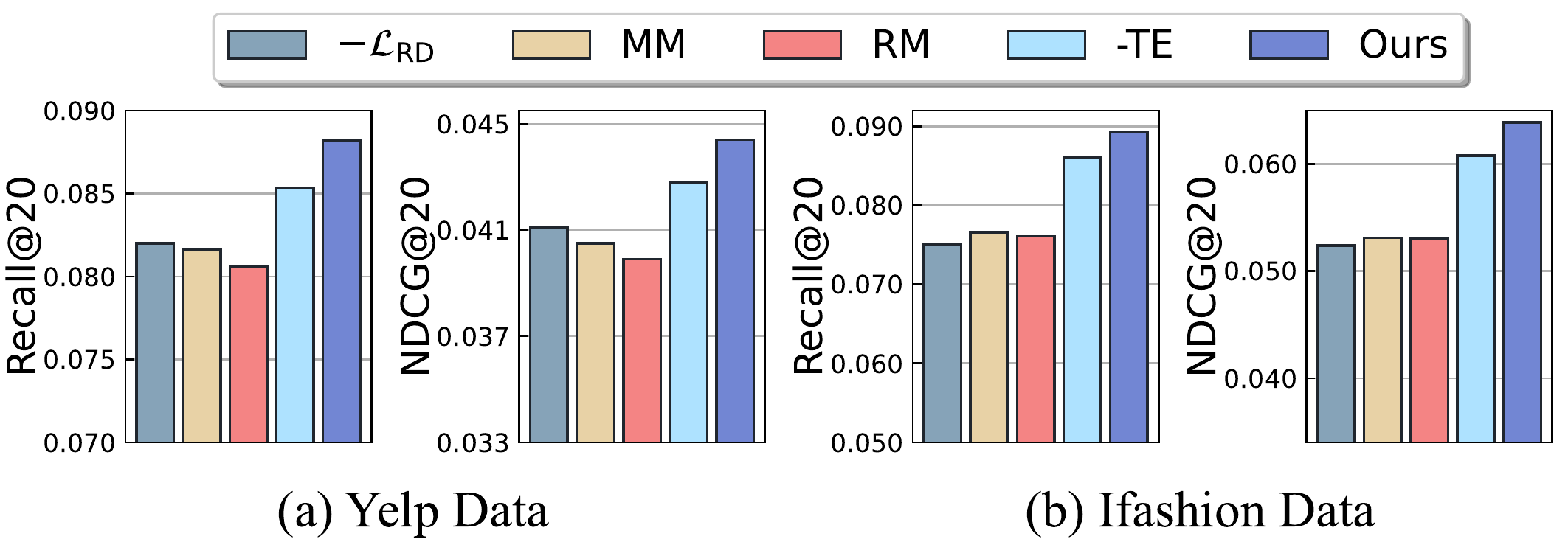}
    \vspace{-0.25in}
    \caption{Performance of ablated models on Yelp and Ifashion datasets in terms of \emph{Recall@20} and \emph{NDCG@20}.}
    \vspace{-0.15in}
    \label{fig:ablation}
\end{figure}

\subsection{Overall Performance Comparison (RQ1)}

We report the overall performance of \model\ and all compared baselines in terms of \emph{Recall@K} and \emph{NDCG@K} under top-10, top-20, and top-40 settings in Table~\ref{tab:overall}. To validate the superiority of our model compared to the strongest baselines, we conduct the test of significance where \(p\)-val\( < 0.05\) suggests a statistically significant improvement achieved by \model. From the experimental results in Table~\ref{tab:overall}, we mainly have the following observations:

\begin{itemize}[leftmargin=*]

\item \model\ consistently outperformed all baselines, including the strong SSL-enhanced methods (\eg, SGL, HCCF) by a large margin. We ascribe this superiority to our rationale-aware SSL augmentation, which automatically derived informative self-supervision signals from the learned collaborative rationale. In contrast, models with stochastic SSL augmentations (\eg, Dropout in SGL) performed much worse due to the possible loss of important graph structural information of sparse users and long-tail items in their randomized contrastive views. Moreover, although HCCF and NCL both adopted carefully-designed hand-crafted CL tasks, they may not be able to provide accurate self-supervision signals (\eg, hard augmented instance) compared to \model. In the face of interaction noise, their pretext tasks are easily misguided by the noisy information contained in the augmented data. Our \model\ tackled these limitations of existing CL models by introducing rationale-aware self-augmentation via masked autoencoding and thus achieved better performance. \\\vspace{-0.12in}


\item Despite the disadvantages of existing CL methods mentioned above, we observed that baseline models with SSL augmentation generally performed better than those without (\eg, NGCF, LightGCN). This could be due to the labeled data scarcity problem of the recommendation task, while SSL can mitigate this problem by introducing additional self-supervision signals from limited observed interactions. Also, incorporating SSL can also alleviate the over-fitting issue of GNNs for user representations, which is used in most strong baselines, on such sparse data and help learn better embeddings for the recommendation.


\end{itemize}

\begin{figure}[t]
    \centering
    \subfigure[Yelp Dataset]{
    \includegraphics[width=\columnwidth]{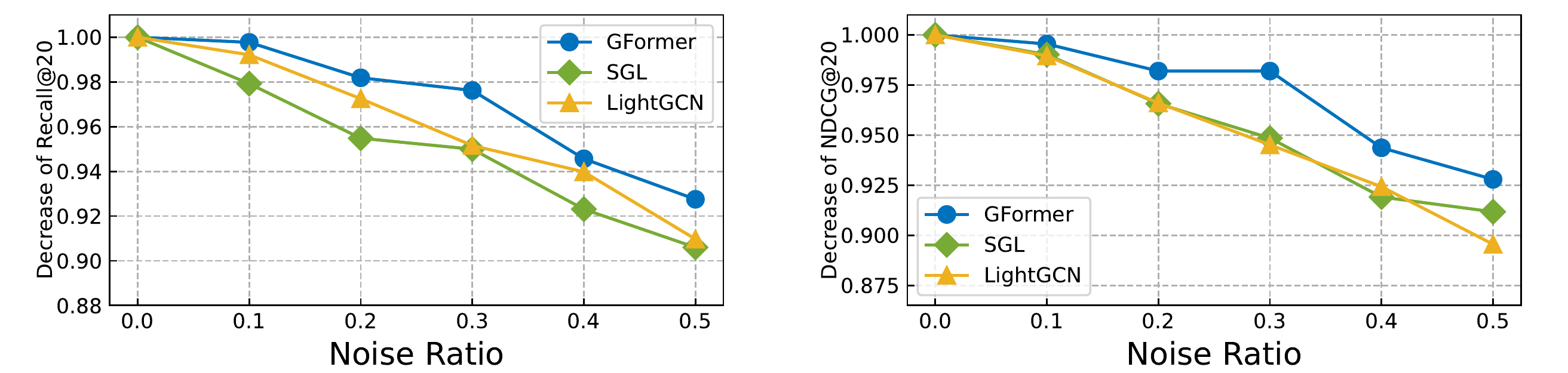}
    }
    \subfigure[LastFM Dataset]{
    \includegraphics[width=\columnwidth]{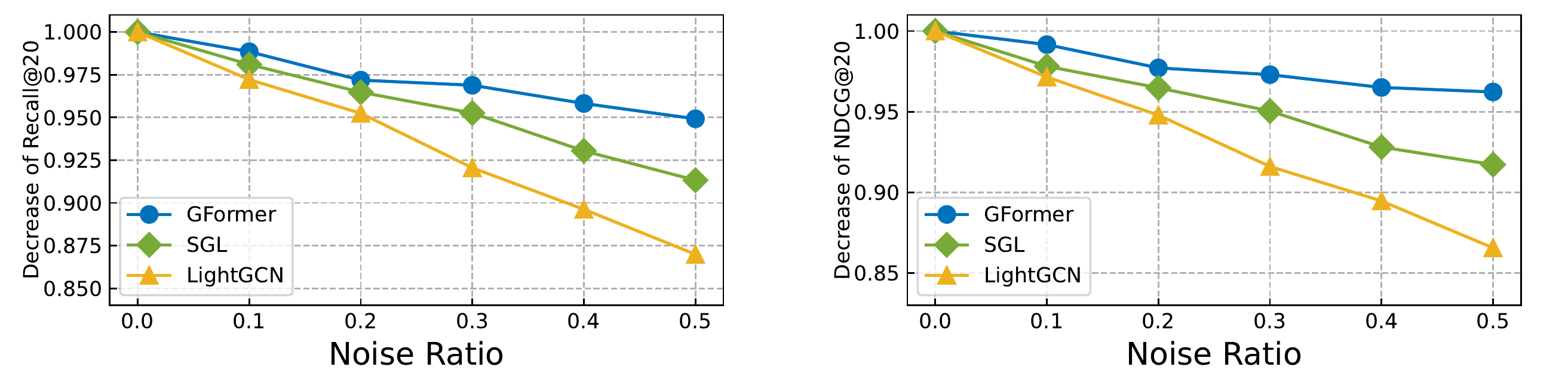}
    }
    \vspace{-0.2in}
    \caption{Performance on Yelp and LastFM datasets with noise perturbation in terms of \emph{Recall@20} and \emph{NDCG@20}.}
    \vspace{-0.1in}
    \label{fig:noise}
\end{figure}

\subsection{Ablation Study (RQ2)}

To study the effectiveness of the key components of \model, we performed an ablation study over several variants, \ie, \textbf{-TE}, \textbf{RM}, \textbf{MM}, and \(\mathbf{-\mathcal{L}}_\textbf{RD}\). Results in terms of \emph{Recall@20} and \emph{NDCG@20} are plotted in Figure \ref{fig:ablation}, from which we had the following discussions.

\subsubsection{\bf Effect of Global Topology Information Injection} In the variant \textbf{-TE}, we replace the topological embedding with pure id-corresponding embedding to disable topology information injection. The results show that the ablated model has a worse performance on both datasets. This is because the injection of global topology information enables our \model\ to capture high-order collaborative relationships for our graph Transformer to encode informative interaction patterns, so that better graph rationale can be provided for both the recommendation task and the reconstruction task to ultimately boost the overall performance.

\subsubsection{\bf Effect of Rationale-Aware Self-Augmentation} To study the influence of the key module of \model, \ie, the rationale-aware self-augmentation module, we replace it with random masking in variant \textbf{RM} and an MLP-based masking strategy in variant \textbf{MM}. Specifically, for an edge \((u,v)\), the \textbf{MM} variant first feeds the embeddings \(\bar{\mathbf{h}}_u,\bar{\mathbf{h}}_v\) into a multi-layer perceptron (MLP) to compute the importance score of nodes \(u, v\), and then obtains the mask probability of the edge by dot product of node importance score, \ie, \(\text{MLP}(\bar{\mathbf{h}}_u) \cdot \text{MLP}(\bar{\mathbf{h}}_u)\). The results show that both variants have a significant performance drop, suggesting that random masking and trivial adaptive masking through MLPs are both unable to discover important graph structures. Instead, they may corrupt informative structures of the interaction graph and introduce additional noise to the SSL task. In contrast, our \model\ avoids these disadvantages by incorporating meaningful self-supervision signals from the learned graph rationale \(R(\mathcal{G})\) and utilizing the complement \(C(R(\mathcal{G}))\) for model denoising, resulting in better performance.

\subsubsection{\bf Effect of Adaptive Rationale Learning} The downstream recommendation loss \(\mathcal{L}_{\text{RD}}\) (Eq.~\ref{eq:l_rd}) plays a crucial role in \model\ by guiding graph invariant rationale learning with adaptive supervision signals. To verify its contribution, we remove \(\mathcal{L}_{\text{RD}}\) in the variant \(\mathbf{-\mathcal{L}}_\text{RD}\), which leads to severe performance degradation. This is because \(\mathcal{L}_{\text{RD}}\) allows our \model\ to discover relevant graph rationale that captures graph structures specifically useful for the recommendation task. Different SSL modules in \model\ can then be optimized in a better-aligned way. Furthermore, we observe that the degradation caused by removing \(\mathcal{L}_\text{RD}\) is larger on the Ifashion dataset. This may be due to the larger amount and denser interaction data of Ifashion compared to Yelp, which provides the loss \(\mathcal{L}_\text{RD}\) with more supervision signals and higher importance.

\begin{figure}[t]
    \centering
    \includegraphics[width=\columnwidth]{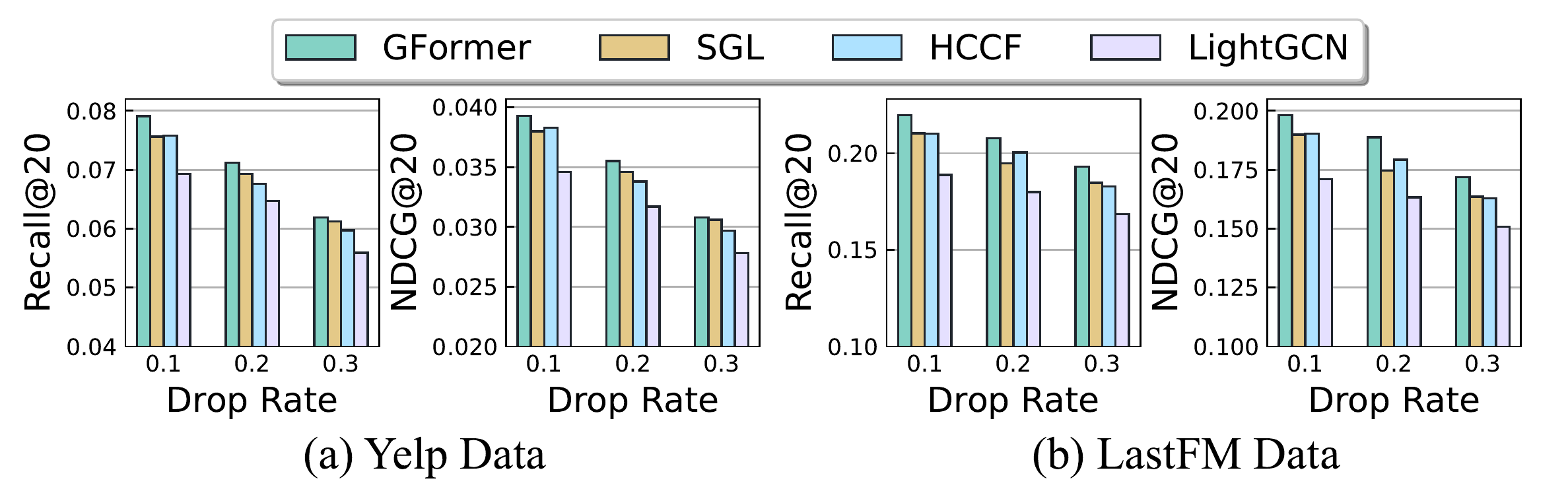}
    \vspace{-0.24in}
    \caption{Performance on Yelp and LastFM datasets under different sparsity level in terms of \emph{Recall@20} and \emph{NDCG@20}.}
    \vspace{-0.1in}
    \label{fig:sparsity}
\end{figure}

\subsection{Model Robustness Study (RQ3)}
In this section, we study the robustness of our model against data noise and data scarcity by testing the model performance on manually damaged training data from the corresponding two dimensions, in comparison to representative baseline methods.
    \subsubsection{\bf Robustness against Data Noise}
    To study the robustness of \model\ against noise perturbation, we randomly inject different proportions (\(10\%, 20\%, 30\%, 40\%, 50\%\)) of edges with artificial noise on the original interaction graph and evaluate the performance of \model\ and representative strong baseline methods on these noisy datasets. As shown by the results illustrated in Figure \ref{fig:noise}, our \model\ consistently achieves the lowest performance degradation under all noise levels. Under lower levels of noise (\eg, \(10\%\) to \(40\%\)), we observe that SSL-enhanced methods (SGL) have a degradation ratio similar to LightGCN, suggesting that stochastic SSL augmentation does not significantly improve robustness against data noise. In contrast, our \model\ adopts an automated SSL paradigm that is rationale-aware, making it possible to discover latent informative structures in a noisy dataset for representation.

    \subsubsection{\bf Robustness against Data Sparsity} 
    We also conduct experiments to evaluate the performance of \model\ on various sparsity levels. Specifically, we drop a certain proportion (\(10\%, 20\%, 30\%\)) of interactions in the dataset and run \model\ as well as representative baseline methods on the sparsified datasets. As shown by the results in Figure \ref{fig:sparsity}, our \model\ outperforms all other models under all sparsity levels, suggesting that our rationale-enhanced SSL framework enables \model\ to generate more meaningful self-supervision signals than common SSL models on sparse data, thereby increasing the model's robustness against data sparsity. Additionally, we observe that the performance degradation with respect to the drop rate is more significant on the Yelp dataset compared to the LastFM dataset. This may be caused by the relatively higher interaction sparsity of the Yelp dataset, such that dropping more interactions severely hinders effective CF modeling.

\begin{figure}[t]
    \centering
    \includegraphics[width=\columnwidth]{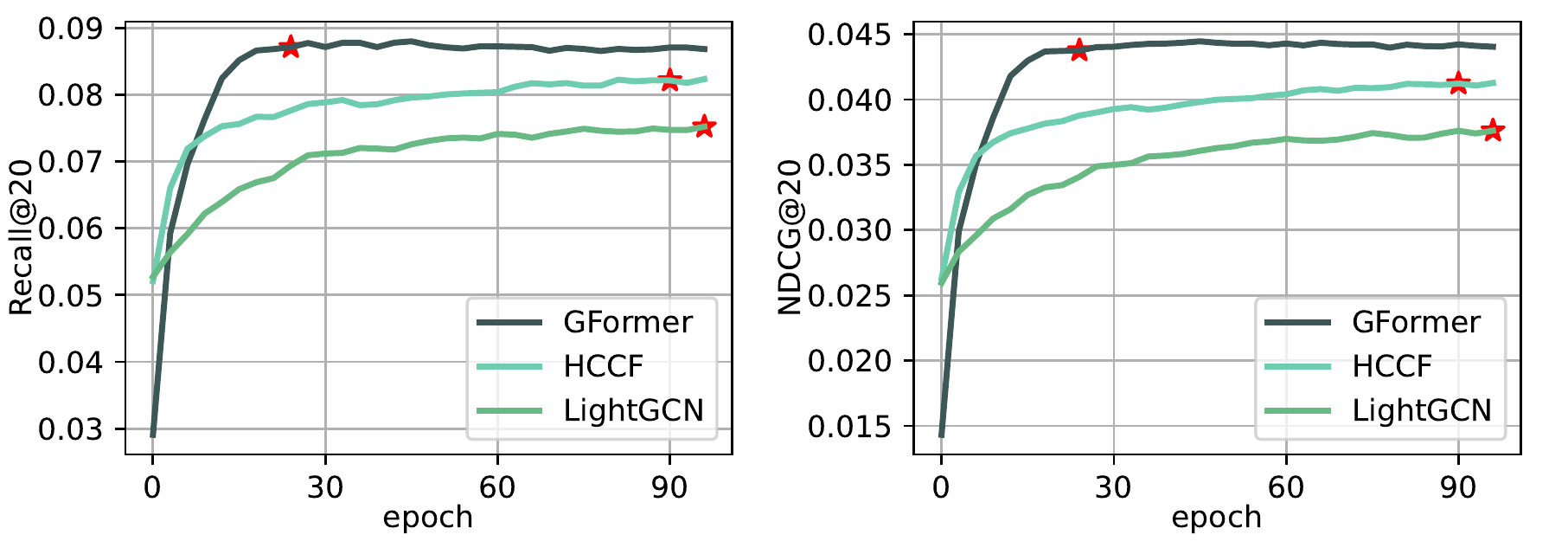}
    \vspace{-0.25in}
    \caption{Test results in terms of \emph{Recall@20} and \emph{NDCG@20} \wrt\ training epochs on Yelp dataset. The stars represent points of convergence. Compared with baselines, the faster convergence rate of our \model\ can be observed.}
    \vspace{-0.1in}
    \label{fig:convergence}
\end{figure}

\subsection{Model Convergence Study (RQ4)}
To analyze the training efficiency of our proposed \model, we plot the convergence curve (\ie, test metric values with respect to training epochs) of \model\ and three strong baseline methods, including SSL-enhanced methods (\eg, HCCF), in Figure \ref{fig:convergence}. It can be observed from the stars indicating convergence that \model\ only takes fewer than 30 epochs to converge, which is much faster compared to other models. We attribute this superiority to the helpful task-adaptive self-supervision signals derived from our learned graph rationale. Stochastic data augmentation strategies cannot actively discover important graph structures for the recommendation task and may provide misleading self-supervision signals during the training stage, thus the baseline models are naturally optimized in a slower manner. Our proposed SSL augmentation focuses on the recommendation task and adapts fast, enabling \model\ to achieve the best performance in an early stage. The faster performance improvement of SSL-based methods compared to LightGCN is likely caused by the low-temperature contrastive learning.

\begin{figure}[t]
    \centering
    \includegraphics[width=0.9\columnwidth]{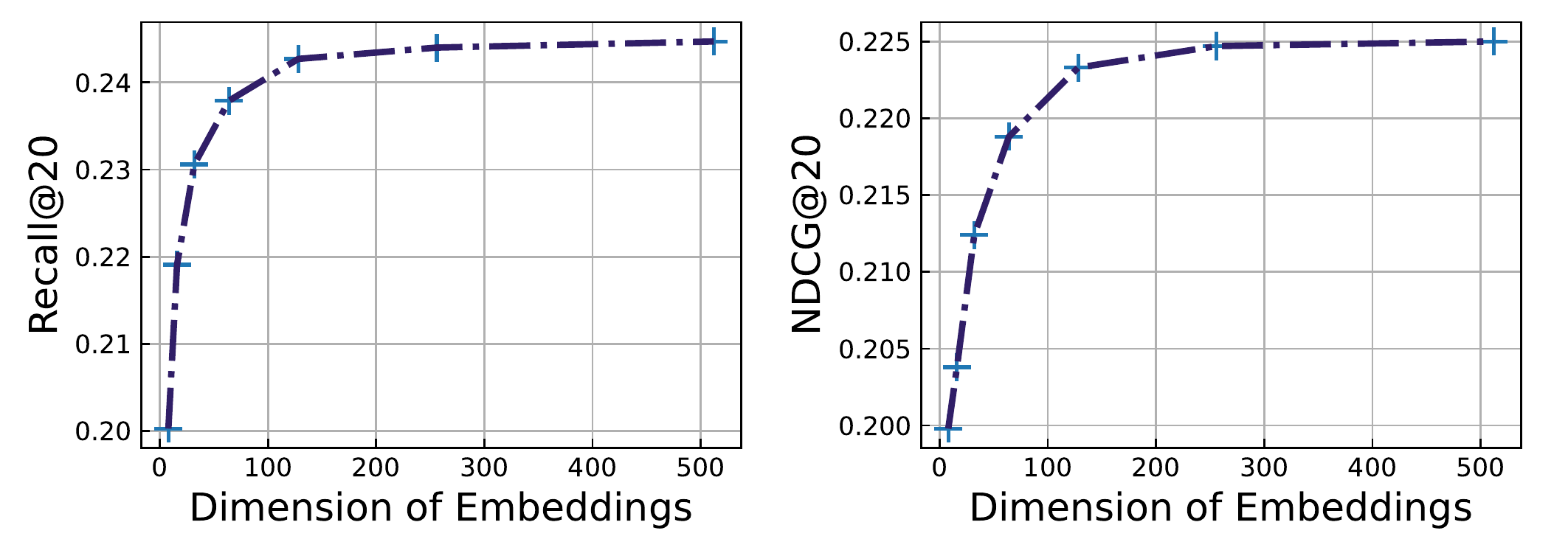}
    \vspace{-0.15in}
    \caption{Performance in terms of \emph{Recall@20} and \emph{NDCG@20} under different embedding dimensionality.}
    \vspace{-0.1in}
    \label{fig:hyp_emb}
\end{figure}

\begin{figure}[t]
    \centering
    \includegraphics[width=\columnwidth]{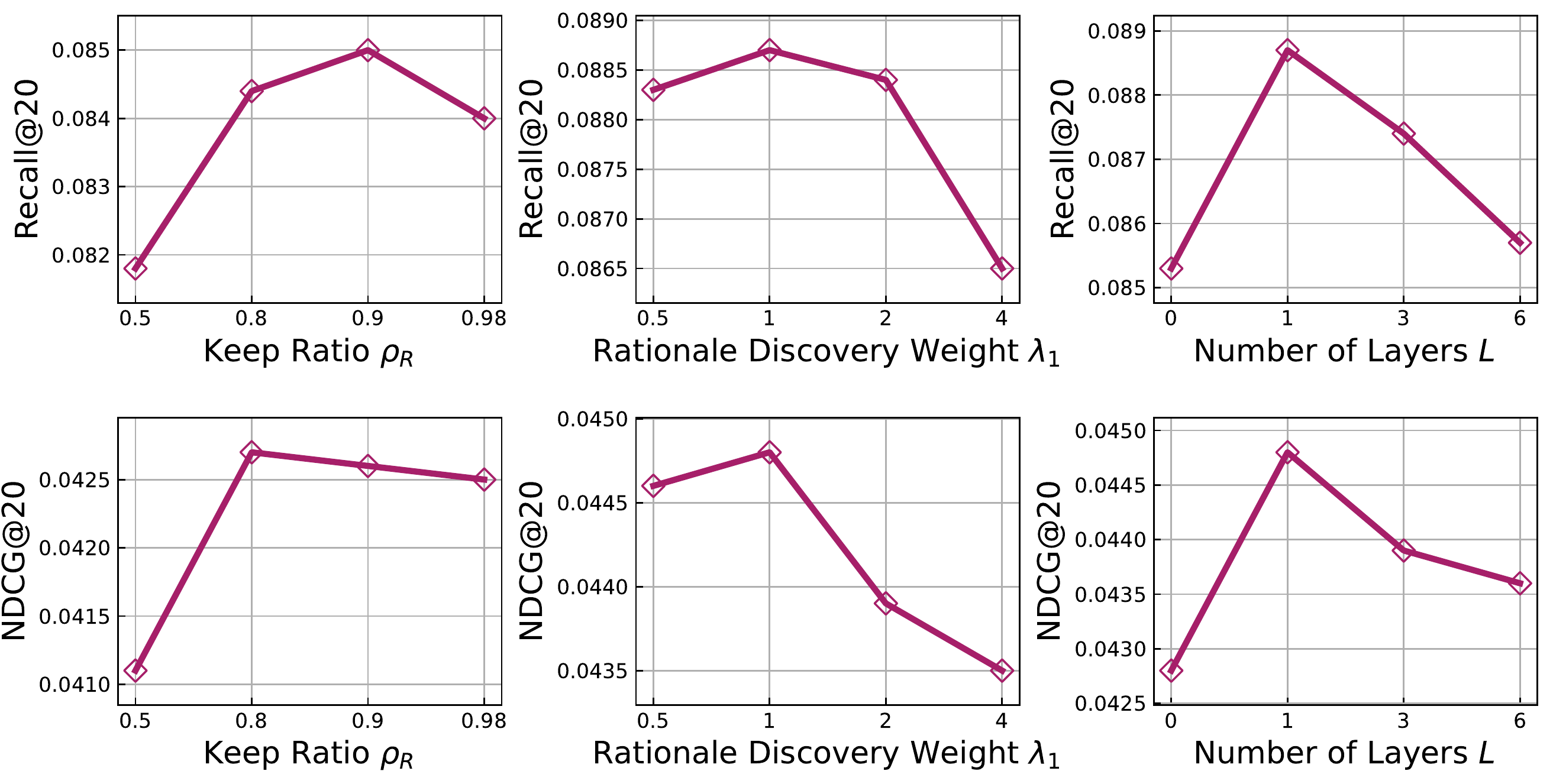}
    \vspace{-0.25in}
    \caption{Performance under different hyperparameter settings in terms of \emph{Recall@20} and \emph{NDCG@20} on Yelp dataset.}
    \vspace{-0.2in}
    \label{fig:hyp}
\end{figure}

\subsection{Hyperparameter Study (RQ5)}
To study the effect of various hyperparameters on the model performance, we present experiment results in terms of \emph{Recall@20} and \emph{NDCG@20} on the LastFM dataset in Figure \ref{fig:hyp_emb} and \ref{fig:hyp}. The following observations corresponding to different hyperparameters are made:
\begin{itemize}[leftmargin=*]

    \item \textbf{Dimension of latent space \(d\)}: We tune the size of user and item embeddings from 8 to 512. As shown in Figure \ref{fig:hyp_emb}, an increase in the embedding dimension leads to significant performance improvement at the beginning. This is because a larger dimensionality allows our topological embedding to capture richer information about the global user-item topology and provide useful representations for other modules in \model. However, too large sizes (\eg, \(d>256\)) only bring subtle improvement, since this may cause the over-fitting problem of GNNs. \\\vspace{-0.12in}

    \item \textbf{Keep ratio \(\rho_R\)}: This hyperparameter controls the proportion of interaction edges to be selected and form the graph rationale \(\mathcal{G}_R\). Results show that a low keep ratio harms model performance since insufficient collaborative relations are obtained in the graph rationale for representation learning. However, setting the keep ratio close to 1 also leads to a performance drop, because noisy interactions in the original graph cannot be adequately dropped.  \\\vspace{-0.12in}

    \item \textbf{Collaborative rationale discovery weight \(\lambda_1\)}: This hyperparameter controls the regularization strength of the objective from the downstream recommendation task. \(\mathcal{L}_{\text{RD}}\) is used to guide the graph rationale discovery in our masked graph transformer paradigm. From the results, we observe that applying large enough weights greatly improves the model performance due to the benefits brought by the discovery of task-relevant graph rationale for augmentation. However, setting the weight for \(\mathcal{L}_{\text{RD}}\) too large may be counterproductive due to the overfitting effect. \\\vspace{-0.12in}




    \item \textbf{Number of layers \(L\) in Global Information Injection}: In our \model, the designed global information injection module can achieve competitive results with one iteration. We found in our experiments that global information encoding with too many layers results in serious damage to performance. Such over-smoothed representations may adversely affect the discovery of rationale-aware augmentation with indistinguishable embeddings. Additionally, setting \(L=0\), \ie\ removing global topology information injection (Section 3.1.2), leads to even worse performance due to insufficient modeling of high-order relations. \\\vspace{-0.12in}

\end{itemize} 

\begin{figure}[t]
    \centering
    \includegraphics[width=\columnwidth]{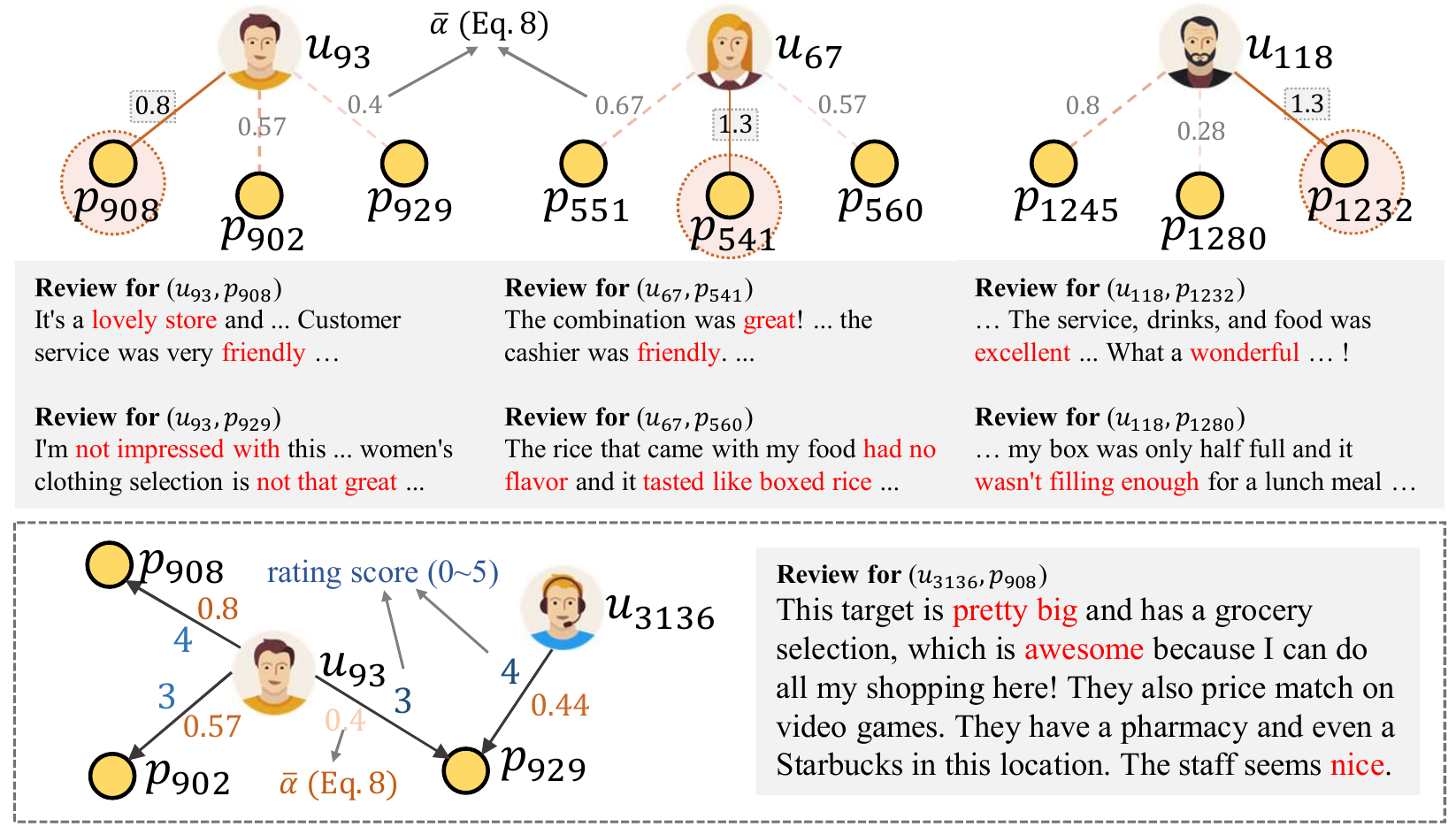}
    \vspace{-0.15in}
    \caption{Case study of collaborative rationale discovery for distilling informative knowledge from noisy interactions.}
    \vspace{-0.05in}
    \label{fig:case}
\end{figure}

\subsection{Case Study (RQ6)}
To study the interpretation ability of \model, a case study is performed over several representative users sampled from the Yelp dataset. Specifically, we inspect the corresponding reviews and ratings given by a user to their interacted items, which are not used for model training, and see whether they match our correlation score \(\bar{\alpha}\) for collaborative rationale discovery.
As illustrated in Figure \ref{fig:case}, the items with the highest correlation scores for users \(u_{93},u_{67},u_{118}\) all match the items they gave positive feedback, while edges to items with unsatisfying reviews are generally given lower scores. These results suggest that \model\ can emphasize learning on informative user-item correlations and utilize these interaction data to enhance the model learning through the reconstruction SSL.
Furthermore, when we investigate the specific ratings given by \(u_{93}\), we observe that the edge with the highest correlation score (\ie, \((u_{93},p_{908})\)) corresponds to the user's highest rating. Meanwhile, although \(u_{3136}\) gives a higher rating (4) than \(u_{93}\) (3) to item \(p_{929}\), the edge \((u_{3136}, p_{929})\) is given a low correlation score similar to \((u_{93}, p_{929})\). This could be due to the fact that user \(u_{3136}\) is very likely to provide many high ratings (36 out of 47 of all ratings given by \(u_{3136}\) are 4 or 5). Our rationale discovery module successfully identifies such bias and adaptively adjusts the correlation score with respect to different user behaviors.

%% file: conclusion.tex
\section{Conclusion}
\label{sec:conclusion}

This paper aims to uncover useful user-item interaction patterns as rationales for augmenting collaborative filtering with the learning of invariant rationales for SSL. Our proposed \model\ model provides guidance to distill semantically informative graph connections with the integration of global topology embedding and task-adaptation. Our work opens avenues for constructing rationale-aware general augmentation through masked graph autoencoding. Our empirical results suggest that SSL-based augmentation with effective rationalization can facilitate user preference learning, and thus significantly boost recommendation performance. While our new model already endows adaptive augmentation with task-aware rationale discovery, it is an interesting open question on how to adapt it to other recommendation scenarios, such as social-aware recommendation and knowledge graph-enhanced recommenders.